\documentclass[aps,twocolumn,superscriptaddress,nobibnotes,letterpaper]{revtex4-2}

\usepackage[pdftex]{graphicx}
\usepackage[pdftex]{epsfig}
\usepackage{amsmath}
\usepackage{amssymb}
\usepackage{amsfonts}
\usepackage{color}
\usepackage{wrapfig}
\usepackage{eucal}
\usepackage{hhline}
\usepackage{threeparttable}
\usepackage{supertabular}
\usepackage{multirow}
\usepackage{tabularx}
\usepackage{siunitx}
\usepackage{float}
\usepackage{sidecap}

\usepackage{soul}

\usepackage{array}
\usepackage[colorlinks=true, allcolors=blue]{hyperref}

\newcommand{\und}[1]{_\textrm{#1}}

\definecolor{cgreen}{rgb}{.1,.6,.1}
\definecolor{co}{rgb}{.1,.6,.6}
\definecolor{orange}{rgb}{.9,.4,.0}

\newcolumntype{C}[1]{>{\centering\arraybackslash}p{#1}}

\begin{document}
\pagenumbering{arabic}

\title{Ultra-low-noise Microwave to Optics Conversion in Gallium Phosphide}\thanks{This work was published in \href{https://doi.org/10.1038/s41467-022-34338-x}{Nature Commun.\ \textbf{13}, 6583 (2022).}}

\author{Robert Stockill}\thanks{These authors contributed equally to this work.}
\affiliation{Kavli Institute of Nanoscience, Department of Quantum Nanoscience, Delft University of Technology, 2628CJ Delft, The Netherlands}
\affiliation{QphoX B.V., 2628XG Delft, The Netherlands}

\author{Moritz Forsch}\thanks{These authors contributed equally to this work.}
\affiliation{Kavli Institute of Nanoscience, Department of Quantum Nanoscience, Delft University of Technology, 2628CJ Delft, The Netherlands}

\author{Frederick Hijazi}
\affiliation{Kavli Institute of Nanoscience, Department of Quantum Nanoscience, Delft University of Technology, 2628CJ Delft, The Netherlands}
\affiliation{QphoX B.V., 2628XG Delft, The Netherlands}

\author{Gr\'{e}goire Beaudoin}
\affiliation{Centre de Nanosciences et de Nanotechnologies, CNRS, Universit\'{e} Paris-Saclay, C2N, 91767 Palaiseau, France}

\author{Konstantinos Pantzas}
\affiliation{Centre de Nanosciences et de Nanotechnologies, CNRS, Universit\'{e} Paris-Saclay, C2N, 91767 Palaiseau, France}

\author{Isabelle Sagnes}
\affiliation{Centre de Nanosciences et de Nanotechnologies, CNRS, Universit\'{e} Paris-Saclay, C2N, 91767 Palaiseau, France}

\author{R\'{e}my Braive}
\affiliation{Centre de Nanosciences et de Nanotechnologies, CNRS, Universit\'{e} Paris-Saclay, C2N, 91767 Palaiseau, France}
\affiliation{Universit\'{e} Paris-Cit\'{e}, 75006 Paris, France}
\affiliation{Institut Universitaire de France (IUF), France}

\author{Simon Gr\"oblacher}
\email{s.groeblacher@tudelft.nl}
\affiliation{Kavli Institute of Nanoscience, Department of Quantum Nanoscience, Delft University of Technology, 2628CJ Delft, The Netherlands}
\affiliation{QphoX B.V., 2628XG Delft, The Netherlands}

\begin{abstract}
Mechanical resonators can act as excellent intermediaries to interface single photons in the microwave and optical domains due to their high quality factors. Nevertheless, the optical pump required to overcome the large energy difference between the frequencies can add significant noise to the transduced signal. Here we exploit the remarkable properties of thin-film gallium phosphide to demonstrate bi-directional on-chip conversion between microwave and optical frequencies, realized by piezoelectric actuation of a Gigahertz-frequency optomechanical resonator. The large optomechanical coupling and the suppression of two-photon absorption in the material allows us to operate the device at optomechanical cooperativities greatly exceeding one. Alternatively, when using a pulsed upconversion pump, we demonstrate that we induce less than one thermal noise phonon. We include a high-impedance on-chip matching resonator to mediate the mechanical load with the 50-\si{\ohm} source. Our results establish gallium phosphide as a versatile platform for ultra-low-noise conversion of photons between microwave and optical frequencies.
\end{abstract}

\maketitle

\section*{Introduction}

\begin{figure}
\includegraphics[width=1\columnwidth]{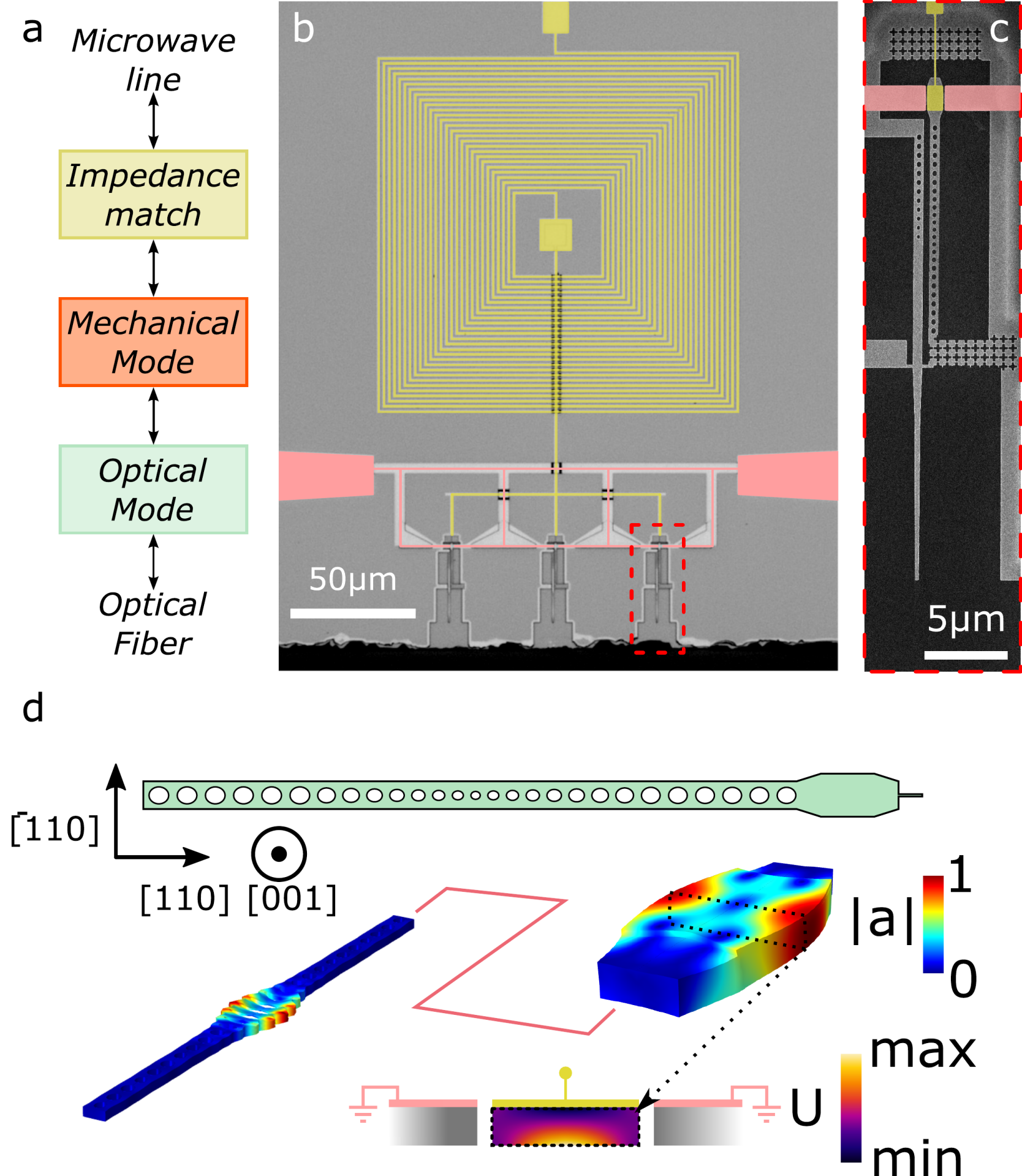}
\caption{\textbf{Conversion Device.} a) Schematic showing the mechanically-mediated conversion process between microwave and optical frequencies. b) Microscope image of the full device consisting of a spiral inductor (top) and three nanobeam OMCs (bottom). Only one of the OMCs is used at any one time. The image is overlaid with false coloring to clarify the ground (red) and center conductor (yellow). c) Zoom-in to one of the nanobeam OMCs. The optical cavity in the nanobeam is evanescently coupled to the optical waveguide, to which we couple with a lensed fiber. The nanobeam is mechanically connected to the piezoelectric resonator and the whole structure is secured to the environment through acoustic shielding. d) Illustration of the combined electro-opto-mechanical converter and the relevant coordinate system (top). Simulated breathing modes of the OMC (left) and piezoelectric resonator (right) around 2.8~\si{\giga\hertz}. The breathing mode is coupled to an out-of-plane electric potential gradient (bottom, center block, see Methods), which we induce with the coplanar electrodes.}
\label{figure1}
\end{figure}

The physical carrier of quantum information plays a crucial role in how the information is processed, communicated and measured. Encoding quantum information in microwave-frequency photons has allowed for the development of circuit~\cite{Arute2019} and spin-based~\cite{Veldhorst2015,Hendrickx2021} quantum information processing. At the same time, optical photons at telecom wavelength are a natural carrier for quantum information~\cite{Yu2020a}, benefitting from particularly low loss rates in optical fibers~\cite{Nagayama2002}, an effectively noise-free environment at room temperature and near-unity efficiency single-photon detection technology~\cite{Chang2021}.

Bridging the five-orders-of-magnitude frequency difference between the microwave and optical domains will allow for flexible manipulation and communication of quantum information~\cite{Zeuthen2020,Zhong2020,Krastanov2021}, enabling both networked quantum computation and complex processing of long-range entangled states.  Steady progress in various experimental realizations of such microwave-to-optics converters has been achieved in the past few years~\cite{Lauk2020}, with significant challenges remaining. These include low-noise operation, efficient transduction, and the construction of scalable platforms. In particular, the requirement of an optical-frequency pump to compensate for the large frequency mismatch introduces the possibility of absorption-based noise which can overwhelm the ultra-low power signal and corrupt the quantum information.

Owing to weak electro-optic coupling rates, the large optical pump powers required for efficient operation typically prohibit direct conversion without the addition of significant absorption-induced noise~\cite{Hease2020}. An alternative is to find intermediaries which can reduce the required optical pump size, while at the same time providing high-cooperativity interfaces to both electrical and optical domains, a role for which mechanical resonators are an attractive system~\cite{Bagci2014,Balram2016,higginbotham2018,Han2020,Jiang2020,Arnold2020,Peairs2020}. To this end, the interaction between a mechanical oscillator and a co-localized optical mode has enabled the creation and measurement of nonclassical states of motion~\cite{Riedinger2016,Riedinger2018,Wallucks2020}. At the same time, by using the piezoelectric interaction or mechanical modification of resonator capacitances mechanical modes have been successfully interfaced with excitations in superconducting qubits~\cite{OConnell2010,Chu2018}. A recent demonstration has shown the generation of optical photons from a superconducting qubit embedded within a transducer~\cite{Mirhosseini2020}, however the close proximity of the transducer to the qubit resulted in the production of quasi-particles in the superconductor, disturbing the qubit coherence. The realisation of a stand-alone quantum transducer, which could in turn be connected to a shielded quantum system, remains an ongoing challenge.

Gallium phosphide (GaP) combines a high refractive index ($n\und{GaP} = 3.1$) and a large band gap (2.3~eV) with a non-centro-symmetric crystal structure enabling applications for low-loss integrated photonics in the telecom band~\cite{Wilson2020}. Thanks to the suppressed two-photon absorption of telecom-wavelength light and the intrinsic piezoelectric coupling in GaP, the material has shown promise in a demonstration of nonclassical optomechanics~\cite{Stockill2019} as a suitable host for a full microwave-to-optical transducer.

Here we build on the impressive properties of GaP (for more details see the Methods) and realize microwave-to-optics conversion using this material. We enhance our previously demonstrated optics-to-mechanical interface with an electrical-to-mechanical interface, whereby we directly actuate a breathing mode of the optomechanical resonator. We include an impedance matching network to mediate the high-impedance load presented by the mechanical resonance with a 50-\si{\ohm} source~\cite{Wu2020}. The ultra-low absorption of telecom photons in the material enables operate of the device with optomechanical cooperativities $C\und{om} \gg 1$ under continuous driving. For pulsed operation, we demonstrate total conversion efficiencies of $6.8\times10^{-8}$, featuring a mechanical-to-optical efficiency of $1.9\times10^{-2}$ (see Methods), while maintaining a thermal occupation in the nanobeam mechanical mode of only $n\und{th} = 0.55\pm0.05$ phonons.

\begin{figure*}
\includegraphics[width=1.5\columnwidth]{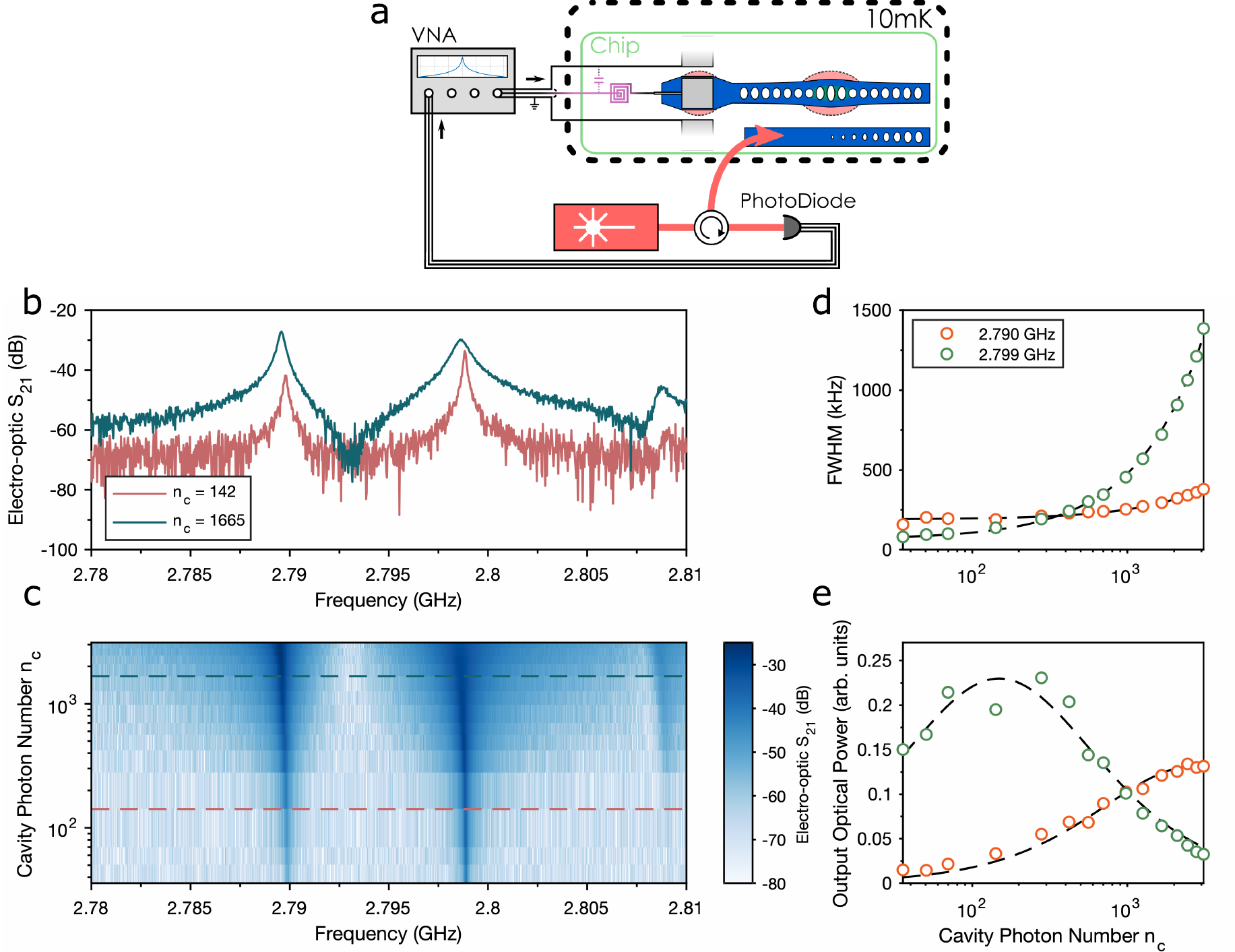}
\caption{\textbf{Device characterisation using an electro-optic S21 measurement.} \textbf{a} Setup for continuous conversion measurement. The device is driven by the output of a vector network analyzer (VNA). The mechanical mode is then read-out using a CW laser which is red-detuned from the optical cavity by $\sim$$\omega\und{m}$. The reflected light is then recorded on a high-frequency photodiode and the signal fed back to the VNA. \textbf{b} Electro-optic conversion signal around 2.8 GHz. Red (blue) curve taken for $n\und{c} = $ 142 (1665). In this electro-optic S21 spectrum, we can see two prominent modes at 2.790~\si{\giga\hertz} and 2.799~\si{\giga\hertz}. \textbf{c} Power-dependent electro-optic conversion signal. \textbf{d} Conversion linewidth for varying optical cavity photon number. Green (orange) points are for the mode at 2.799 (2.790) GHz. \textbf{e} Photon-number-dependent un-normalized optical output power for the same two modes.}
\label{figure2}
\end{figure*}

\section*{Results}

Our converter consists of an optomechanical and an electromechanical interface (Fig. \ref{figure1}a). The former is realized by a nanobeam optomechanical crystal (OMC)~\cite{Eichenfield2009b}, with an optical resonance around 1550~\si{\nano\meter} and a mechanical breathing mode (Fig.~\ref{figure1}d) with a resonance frequency of $\omega\und{m}=2\pi\times$2.81~\si{\giga\hertz}. We fabricate our devices out of a 230-nm-thick layer of suspended gallium phosphide~\cite{Schneider2019}.  The electromechanical interface is realized by a piezoelectric resonator with a breathing mode of the same symmetry and polarisation as the nanobeam mode (Fig.~\ref{figure1}d). These two interfaces are mechanically connected and are brought into resonance by careful design of the width of the piezoelectric resonator. In order to facilitate the mechanical coupling between the two parts of the converter, we design the OMC such that the mechanical mode can leak out of the resonator~\cite{Fang2017}. We attach the converter to its surrounding through an acoustic shield~\cite{Wallucks2020,MacCabe2020} to avoid an increased mechanical decay rate into the environment. A direct benefit of this approach is that we can spatially separate the optical cavity from the metallic electrodes, which avoids perturbations of the optical resonance as well as negative effects on the superconducting electrodes due to a nearby optical cavity. 

The actuation of the piezoelectric resonator's breathing mode is facilitated by alignment of the device axis (Fig.~\ref{figure1}d) with  the $\left[ 110\right]$ axis in the zincblende lattice of gallium phosphide. For this orientation, the breathing mode of the piezoelectric resonator produces an out-of-plane electric field (bottom of Fig.~\ref{figure1}d). We use a coplanar-waveguide electrode configuration to produce this field orientation and actuate the mechanical mode, which results in a simulated piezoelectric coupling strength for the piezo-resonator alone of $k\und{eff}^2 =  1.59\times10^{-6}$ (see Methods). While the small physical size of the piezoelectric resonator is compatible with large optomechanical coupling rates, the resulting small simulated capacitance of $C\und{res} = 0.17$~\si{\femto\farad} results in a large electrical impedance (see Methods, we expect $R\und{m} = 2.1$~\si{\mega\ohm} for a mechanical quality factor of 100,000), such that conversion from a 50~$\Omega$ coaxial line is very inefficient. To overcome this challenge we pattern an electrical impedance-matching circuit on the chip, consisting of a spiral inductor fabricated from a 90-nm-thick layer of molybdenum rhenium (MoRe) (Fig.~\ref{figure1}b)~\cite{Singh2014a}. The capacitance of the impedance matching circuit is limited by the parasitic capacitance of the spiral to ground. In order to match a high-Z impedance, the characteristic impedance of the matching circuit, $Z\und{match}$, should be large, while the resonance frequency should be matched to the mechanical resonance $\omega\und{match} = \omega_m$. For resonant circuits we extract a circuit impedance of $Z\und{match} = 3.1$~\si{\kilo\ohm}, suitable for matching a load of 192~\si{\kilo\ohm} to a 50~\si{\ohm} source.

\subsection*{Characterization}

We fabricate the device shown in Figure~\ref{figure1}b and cool it to Millikelvin temperatures in a dilution refrigerator to initialize the mechanical mode into its ground state. First, we record a reflection-spectrum of the optical cavity by scanning a tunable laser across the resonance. The spectrum reveals a mode at $\lambda\und{c} = $1555.4~\si{\nano\meter} ($\omega\und{c} = 2\pi\times$192.743~\si{\tera\hertz}) with a linewidth (FWHM) of $\kappa = 2\pi\times$4.17~\si{\giga\hertz}, over-coupled to the optical waveguide with external coupling rate  $\kappa\und{e} = 2\pi\times$2.54~\si{\giga\hertz} (see Methods).

We then proceed to characterise the mechanical modes of interest. We identify these modes by performing an electro-optic S21 measurement, where we use a vector network analyzer (VNA) to provide an RF-drive tone to the electrical input of the device which excites the mechanical modes. At the same time, we use a laser stabilized to the red sideband of the optomechanical cavity ($\omega\und{l}=\omega\und{c}-\omega\und{m}$) to upconvert the signal which we monitor in the high-frequency noise of the reflected light using a high-frequency photodetector. The setup is displayed in Figure~\ref{figure2}a The resulting spectrum is shown in Fig.~\ref{figure2}b, where we can identify two efficient modes at 2.790~\si{\giga\hertz} and 2.799~\si{\giga\hertz}. We perform an electrical S11 reflection measurement of the impedance matching resonator, which reveals a resonance at 2.85~\si{\giga\hertz}, 50~\si{\mega\hertz} blue-detuned from the relevant mechanical modes (see Methods). Further, we confirm the bi-directional nature of the transduction process by measuring downconversion from the microwave to optical domain (see Methods).

\subsection*{Continuous conversion}

We obtain insight into the optomechanical device operation by performing conversion while sweeping the input optical power. The electro-optic signal for a range of optical cavity photon numbers ($n\und{c}$) is plotted in Figure~\ref{figure2}c. From fits to the two most efficient frequencies, we extract the transduction bandwidth and un-normalized output optical power, displayed in Figure~\ref{figure2}d and e, respectively. For larger optical-cavity photon numbers we observe a linear increase in the transduction bandwidth for both modes, owing to optomechanical damping. We use this increase in damping rate to extract the single-photon optomechanical coupling rates for the two modes (see Methods), from which we find a value of $g_0 = 2\pi\times 700 \pm 8$~\si{\kilo\hertz} ($2\pi\times272\pm14$~\si{\kilo\hertz}) for the mode at 2.799~\si{\giga\hertz} (2.790~\si{\giga\hertz}). We also extract the intrinsic damping rate of the mechanical modes, $\gamma\und{m}$, from these fits, and find a rate of $\gamma\und{m} = 2\pi\times67\pm8$~\si{\kilo\hertz} ($\gamma\und{m} = 2\pi\times191\pm14$~\si{\kilo\hertz}) for the 2.799~\si{\giga\hertz} (2.790~\si{\giga\hertz}) mode, resulting in a single photon cooperativity (defined as $C\und{0} = 4g\und{0}^2/\kappa\gamma\und{m}$) of $C\und{0} =\left(7.0\pm0.81\right)\times10^{-3}$ ($C\und{0} =\left(3.7\pm0.5\right)\times10^{-4}$).

We plot the evolution of the output optical power in figure \ref{figure2}e. We recover peaked transduction efficiencies with  maximum values for the 2.799~GHz (2.790~GHz) mode at 148 (2720) photons in the optical cavity, corresponding to optomechanical cooperativities of 1.04 (1.01), in good agreement with the expected behaviour $\eta\und{om} \propto C\und{om}/\left(1 + C\und{om}\right)^2$, where $C\und{om} = n\und{c}C_0$. We reach unity cooperativity for an incident optical power of 0.5~$\mu$W (11.3~$\mu$W). While the relatively low electromechanical cooperativity of the device prevents us from directly measuring the absolute electromechanical coupling rate, we can determine that the electromechanical coupling rate for the 2.790~GHz mode is around 3 times larger than the 2.799~GHz mode. This difference suggests that the higher frequency mode is predominantly located in the opto-mechanical resonator, and the lower frequency mode in the piezo-resonator.

We note that we are able to record reproducible efficiency results over this wide range of optomechanical cooperativities thanks to the absence of two-photon absorption induced thermo-optic bistability in GaP~\cite{Schneider2019}. These results also demonstrate the lack of significant damage to the superconducting metal circuitry for these operating powers.

While the measurements shown in Fig.~\ref{figure2} already confirm microwave-to-optics conversion with this device, the continuous operation in this experiment is accompanied by an elevated thermal occupation of the mechanical mode. Therefore, it is important to investigate the performance of this device in the context of the desirable low-noise regime which is crucial for quantum applications. By operating both the microwave drive and the optical interface in a pulsed fashion, we can reduce the thermal noise of the mechanical mode and access the photon-number conversion efficiency. In the following, we will restrict the discussion to the most efficient mode at 2.799~GHz.

\begin{figure}
\includegraphics[width=\columnwidth]{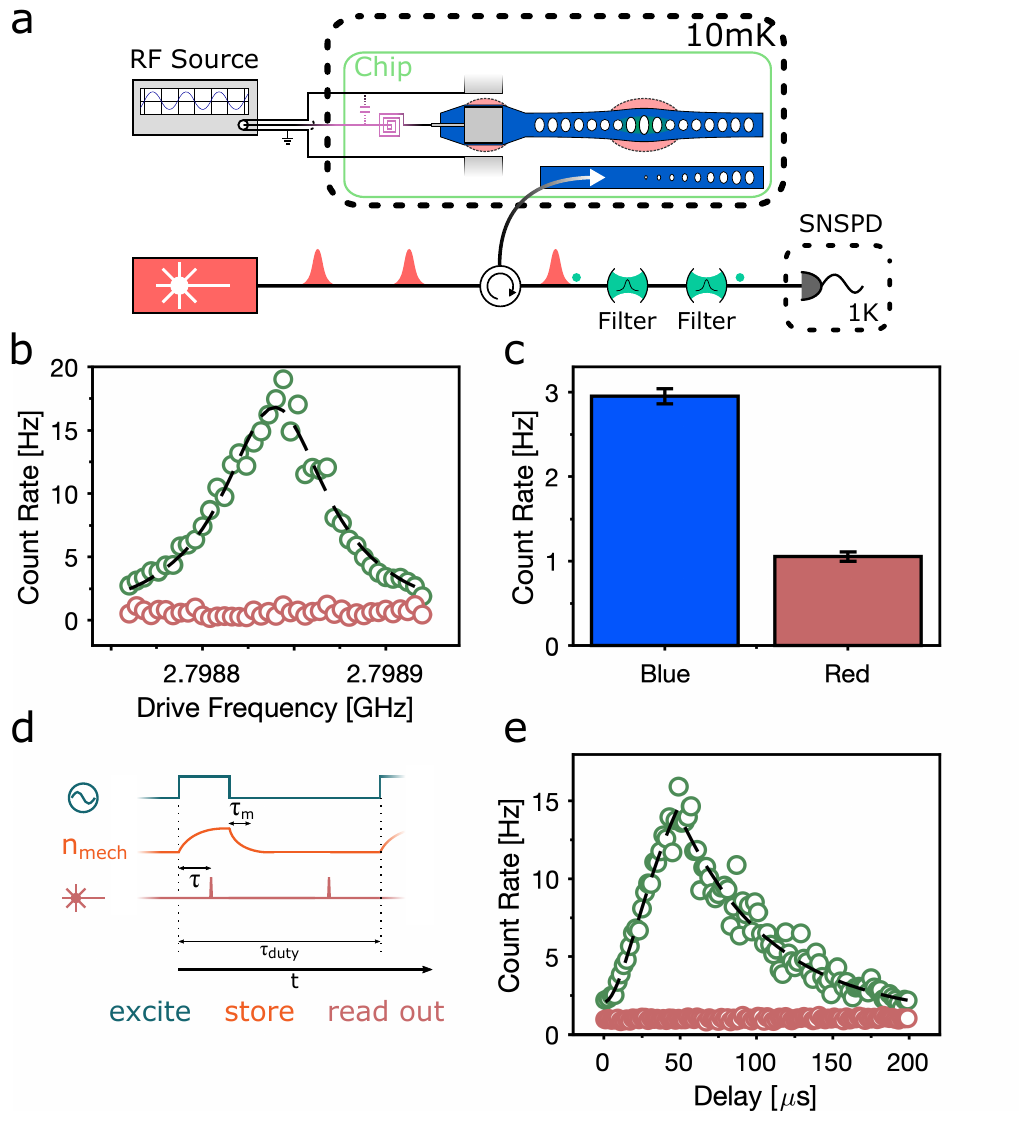}
\caption{\textbf{Pulsed Conversion.} a) Microwave-to-optics conversion measurement setup. We first excite the mechanical mode with a square microwave pulse and then probe the population of the mechanical mode using an optical pulse on the red sideband of the optomechanical cavity ($\omega\und{l}=\omega\und{c}-\omega\und{m}$). We then filter out residual pump light and detect the remaining cavity-resonant photons on SNSPDs. b) Frequency sweep of the pulsed microwave drive across the lowest mechanical resonance. We send a microwave pulse with length 26~\si{\micro\second}, followed by a 40~\si{\nano\second} optical pulse to probe the mode occupation (green datapoints). We let the mode fully re-thermalize again before sending another red-detuned pulse to obtain a base-countrate corresponding to the thermal background (red data points). The green datapoints are overlaid with a Lorentzian fit with a linewidth of $2\pi\times67$~\si{\kilo\hertz}. c) Sideband asymmetry measurement without an RF-drive. Here, we send a sequence of red(blue)-detuned pulses to the device, spaced in time by several mechanical lifetimes. The red (blue) pulses realize the beamsplitter (two-mode squeezing Hamiltonian) and result in count rates $\Gamma\und{R}\propto n\und{th}$($\Gamma\und{B}\propto n\und{th}+1$). Error bars are s.d. d) Schematic illustration of our conversion process. We excite the mode using a square microwave pulse (teal). This pulse leads to a buildup of the mechanical mode's population (orange). After the end of the microwave pulse, the mode population decays with the lifetime of the mode, $\tau\und{m} = 61.4$~\si{\micro\second}. We probe the occupation of the mechanical mode using a red-detuned optical pulse. By setting the delay of this pulse with respect to the beginning of the square pulse, we can trace out the evolution of the mechanical mode's population. We also use a second optical probe pulse after the mechanical mode has decayed in order to measure a reference countrate for an empty mode. e) Time dynamics of the mechanical mode population. Green datapoints show the integrated countrate over the duration of the optical pulse. Dashed curves are fits to the rising and decaying cavity population~\cite{Han2020}.}
\label{figure3}
\end{figure}

\subsection*{Pulsed conversion}

The basic elements of our setup for pulsed operation are displayed in Fig.~\ref{figure3}a. We up-convert the phonons in the mechanical mode to the optical domain with 40-ns long red-detuned optical pulses (at frequency $\omega\und{l}=\omega\und{c}-\omega\und{m}$), separated by sufficient time for the mechancial mode to fully re-thermalize to the mK-environment. These short pulse lengths ensure that we can swap out the mechanical excitation into the optical mode before excess absorption-induced heating commences. We then filter out the upconverted photons from our reflected pump with two Fabry-P\'erot filter cavities in series and detect the converted photons on superconducting nanowire single photon detectors (SNSPDs).

We first find the conversion resonance again by scanning the input microwave frequency and monitoring the rate of single photon clicks, the result displayed in Fig.~\ref{figure3}b, which shows the count rate with (green) and without (red) a 26-\si{\micro\second}-long microwave pulse. We also determine the thermal population of the mode under the same optical pulse energy, by measuring the asymmetry between Stokes and anti-Stokes scattering for blue- and red-detuned laser pulses, respectively. The rates are shown in Figure~\ref{figure3}c, which corresponds to a value of $n\und{th}=0.55\pm 0.05$ for the 40-fJ (314,000-photon) optical pulse energies used here. With separate measurements of the optomechanical coupling rate, we find that for these pulse energies, we convert phonons to optical photons with an efficiency of $p\und{sw} =$~3.2~\%. We note that during the optical pulse the optomechanical cooperativity (defined as $C = 4n\und{ph}g\und{0}^2/\kappa\gamma\und{m}$, where $n\und{ph}$ is the number of photons present in the optical mode due to the pump pulse) exceeds unity, namely $C = 1.74$.

To measure the efficiency of our conversion process in the pulsed scheme, we follow the schedule in Fig.~\ref{figure3}d. We first excite the mechanical mode with a square microwave-frequency pulse. We then upconvert the mechanical occupation to the optical domain with the 40-ns-long optical pulse. We choose this optical pulse length to swap out the mechanical state into the optical mode before excessive heating of the mechanical mode occurs. With a probability $p\und{sw}$ this pulse swaps the mechanical occupation into photons at the cavity resonance frequency, which leaks from the cavity into the coupled waveguide with an outcoupling efficiency of 61~\%. To find the highest efficiency, we scan the optical pulse along the duration of the microwave pulse, the results for a $50$-\si{\micro\second} long microwave pulse are shown in Fig.~\ref{figure3}e. For the mode analysed here, the maximum efficiency occurs for a pulse length of 26~\si{\micro\second} (as displayed in Fig.~\ref{figure3}b). Taking into account external losses including attenuation of microwave lines and additional optical path losses, the latter we measure using the same scheme as in~\cite{Riedinger2016}, we estimate a total conversion efficiency of $6.8\times10^{-8}$. This can be broken up into a power-dependent mechanics-to-optics efficiency of $1.9\times10^{-2}$ ( $=p\und{sw}\kappa_e/\kappa$), and an electrical to mechanical efficiency of $3.6\times10^{-6}$.

The mechanics-to-optics efficiency presented here is linearly dependent on the input optical pump energy, up to the limit of $\kappa\und{e}/\kappa = \eta\und{o} = 0.61$. For smaller energies, the efficiency decreases, however results in a better noise performance. If we operate in a regime where the mechanical-to-optical efficiency is $8\times10^{-3}$, we recover a mode population of $n\und{th} = 0.36 \pm 0.03$ (see Methods). From the single photon cooperativity extracted from the data in figure \ref{figure2} we can estimate the total single-pump-photon microwave-to-optical efficiency for the 2.799 GHz mode of $0.75\times10^{-7}$\cite{Jiang2020}. We note that this is the highest per-optical-photon efficiency recorded for a standalone piezoelectric microwave-to-optics transducer.

The impedance of the piezoelectric interface drops for increasing mechanical quality factor, and it becomes easier to match to the 50~\si{\ohm} source. While we record lifetimes for the mechanical mode of $61.4$~\si{\micro\second}, which set an lower bound on the linewidth of $2\pi\times2.5$~\si{\kilo\hertz} (a quality factor of $1.08\times10^6$), due to mechanical frequency jitter, we recover a minimum linewidth for the mechanical mode of $\gamma\und{m}\sim2\pi\times67$~\si{\kilo\hertz}. This de-phasing effect is also visible in the time dynamics of the loaded mechanical mode we display in figure \ref{figure3}e, where the rise time of $16$~\si{\micro\second} is significantly smaller than the decay time of the mechanical mode. This jitter reduces the efficiency with which we can load the mechanical mode, and raises the time-averaged electrical impedance of the mode. We estimate a reduction in the efficiency of 6.9 for quasi-static noise. Similar increased linewidths have been observed in the mechanical resonances in silicon~\cite{Wallucks2020,MacCabe2020}, and further studies are required to measure the dynamics of this noise source in GaP. Nonetheless, with higher matching circuit impedance, the electrically-enhanced mechanical damping rate would negate these effects. Another key component of the device efficiency is the detuning between the matching resonator and the mechanical mode. At milli-Kelvin temperatures we measure a detuning between the matching network and the mechanical modes of 50-60~\si{\mega\hertz}. When we raise the temperature of the transducer to 4~\si{K} we red-shift the impedance matching resonance through the kinetic inductance of the MoRe film~\cite{Singh2014a}, which compensates for the detuning and increases the efficiency by a factor of 2.2 and 1.7 for the 2.790 and 2.799~\si{\giga\hertz} modes, respectively. An alternative to temperature-tuning of the kinetic inductance is magnetic field tuning, which can compensate for detunings of the order we observe here~\cite{Samkharadze2016,Xu2019}.

One route to higher electromechanical efficiency is through increasing the impedance of the matching circuit by minimizing stray circuit capacitance, which can be achieved through reduction in the size of the impedance matching coil, as well as underetching of the circuit dielectric~\cite{Peruzzo2020}. Through these techniques, values of $ C\und{match}\sim1-2$~\si{fF} are readily achievable, for which we expect to boost the electromechanical efficiency by two orders of magnitude. Alternatively, stronger piezoelectric materials, such as lithium-niobate, used in conjunction with silicon-on-insulator, would allow us to retain the optomechanical coupling rates we measure here, while increasing the electromechanical cooperativity~\cite{Wu2020}. We note that recent demonstrations of mechanically-mediated transducers built entirely from stronger piezoelectric materials, such as all lithium niobate~\cite{Jiang2020} or aluminium nitride~\cite{Han2020}, have achieved higher overall efficiency. Nonetheless, these devices require significantly larger optical pump powers which prohibit efficient operation with sub-phonon thermal excitation.

\section*{Discussion}

Our demonstration of conversion in gallium phosphide shows the material has great promise as the basis of a quantum-capable microwave-to-optics interface. To this end we have demonstrated electromechanical coupling from a 50~\si{\ohm} impedance source to an optically-coupled nanomechanical piezo-resonator through impedance matching. Thanks to the particularly low two-photon absorption in GaP, we operate the device in regimes where the optomechanical cooperativity greatly exceeds one, even in the pulsed regime and while keeping the mechanical oscillator in its ground state. Crucially for quantum transduction, we introduce sub-phonon noise levels during pulsed operation, while maintaining swap efficiencies between phonons and optical photons exceeding 0.05. In future work the electromechanical coupling rate could be increased by using high-impedance electrical resonators~\cite{Samkharadze2016,Peruzzo2020}, and the electromechanical cooperativity could be increased by decoupling the electrical resonator from the 50-\si{\ohm} input. Further significant electromechanical efficiency increase can be expected from including a bottom electrode instead of our coplanar design, as was recently demonstrated in~\cite{Honl2022} for GaP. Already with the current generation of devices we demonstrate a record-high single-pump-photon transfer efficiency for a standalone piezoelectric transducer, thanks to the large optomechanical coupling rate provided by the material.

During preparation of the manuscript, we became aware of related work demonstrating microwave-to-optics conversion in gallium phosphide~\cite{Honl2022}.

\begin{acknowledgments}
	We would like to thank Mario Gely, Bas Hensen, Marios Kounalakis, Igor Marinkovi\'{c}, Sarwan Peiter, Felix Schmidt, Kartik Srinivasan and Andreas Wallucks for valuable discussions and support. We also acknowledge assistance from the Kavli Nanolab Delft. This work is supported by the French RENATECH network (I.S.,R.B.), as well as by the Foundation for Fundamental Research on Matter (FOM) Projectruimte grant (16PR1054) (S.G.), the European Research Council (ERC StG Strong-Q, 676842 and ERC CoG Q-ECHOS, 101001005) (S.G.), and by the Netherlands Organization for Scientific Research (NWO/OCW) (S.G.), as part of the Frontiers of Nanoscience program, as well as through Vidi (680-47-541/994) (S.G.) and Vrij Programma (680-92-18-04) (S.G.) grants. R.S.\ also acknowledges funding from the European Union under a Marie Sk\l{}odowska-Curie COFUND fellowship.
\end{acknowledgments}

\textbf{Author Contributions:}\ R.S., M.F., and S.G.\ planned the experiment and performed the device design. G.B., K.P., I.S.\ and R.B.\ supplied the material. R.S.\ and M.F.\ fabricated the sample and R.S.\, M.F.\. and F.H. performed sample characterization. R.S.\ and M.F.\ performed the measurements, while R.S., M.F.\ and S.G. analyzed the data and wrote the manuscript with input from all authors. S.G.\ supervised the project.

\textbf{Competing Interests:}\ R.S., F.H.\ and S.G.\ declare a financial interest in QphoX B.V. The remaining authors declare no other competing interests.

\textbf{Data Availability:}\ Source data for the plots are available on \href{https://doi.org/10.5281/zenodo.7191940}{Zenodo}.

\setcounter{figure}{0}
\renewcommand{\thefigure}{S\arabic{figure}}
\setcounter{equation}{0}
\renewcommand{\theequation}{S\arabic{equation}}

\clearpage

\section*{Supplementary Information}

\subsection{Piezoelectric properties and orientation}
Gallium phosphide is a III/V semiconductor that crystallizes in the cubic Zincblende lattice. Its piezoelectric properties arise from the small anisotropy given by the two different basis atoms: gallium and phosphorus. For the most common substrate orientation $\left\langle001\right\rangle$ ($\parallel\vec{z}$), the piezoelectric tensor describing these properties is given by:
\begin{equation}
\label{equation_piezo_tensor}
e'_{jk}(\phi)=\frac{e_{14}}{2}
\begin{bmatrix}
0 & 0 & 0 & 2a_2 & -2b_2 & 0 \\ 
0 & 0 & 0 & 2b_2 & 2a_2 & 0 \\ 
-b_2 & b_2 & 0 & 0 & 0 & 2a_2 \\ 
\end{bmatrix},
\end{equation}
where $\phi$ is the angle between the x-axis and the $\left\langle110\right\rangle$ direction, $a_2=sin(2\phi)$ and $b_2=cos(2\phi)$~\cite{Soderkvist1994}. For GaP, the value of $e_{14}$ is $-0.1$\si{\coulomb\per\centi\meter\squared}~\cite{Nelson1968}. Importantly, orienting our structures along the $\left[110\right]$ ($\phi$=0) axis results in non-zero values for $e'_{31}$ and $e'_{32}$, which couple an out-of-plane electric field to X and Y strain, respectively, which allow us to realize this piezoelectrically active breathing mode of the block resonator. We access this vertical electric field through the coplanar electrodes deposited on top of the block, which result in the voltage distribution simulated in Figure~\ref{figureSI1}.

\subsection{Device fabrication}

\textbf{Material Growth}. The starting material for these devices is an epitaxial structure grown by MOCVD on a GaP $\langle 100\rangle$ substrate, consisting of a 1~$\mu$m thick sacrificial layer of Al$_{0.64}$Ga$_{0.36}$P, followed by a 230~nm thick device layer of GaP. The growth was performed in a Veeco Turbodisc D180 reactor under hydrogen as carrier gas, trimethylgallium and trimethylaluminum as organometallic precursors and under phosphine at a reactor pressure of 70~Torr.

The full device fabrication consists of two main parts: first, the patterning of the electrodes and the impedance matching circuit and second, of the optomechanical structures.

\textbf{Metal Layer}. For the first part, we begin by spin-coating AR-P 6200-09 electron beam resist at 40000~\si{rpm} and bake it for 3 minutes at 160~\si{\celsius}, followed by exposing  the bond pads, ground planes, fine electrodes next to the piezoelectric resonator, and several sets of coarse and fine alignment markers for subsequent lithography steps. We then develop the exposed resist by immersion for 1 minute in Pentyl Acetate, followed by a 5~s immersion in Xylene and a final 1 minute IPA rinse. We then sputter $\sim$40~\si{\nano\meter} of molybdenum-rhenium (MoRe) on the developed resist, followed by a lift-off step in an 80~\si{\celsius} bath of Anisole and gentle sonication. After the lift-off is finished, we place the sample in an IPA beaker and sonicate for 5 minutes, followed by blow drying.

We then spin-coat the sample with PMGI-SF11 electron-beam resist at 4000~\si{rpm} and bake it for 5 minutes at 190~\si{\celsius}, followed by exposure of the scaffolding structures, which will later support the air bridges of the LC resonator. The exposed resist is then developed by immersion for 10~s in MF321 developer, 15~s H$\und{2}$O, and 10~s IPA. Next, we re-flow the remaining resist by placing the sample on a pre-heated hotplate at 210~\si{\celsius} for 5 minutes.

On top of the re-flowed PMGI resist, we spin-coat AR-P 6200-18 electron-beam resist at 3250~\si{rpm} and bake at 160~\si{\celsius} for 3 minutes, followed by exposure of the spiral inductor. The resist is again developed by immersion for 1 minute in Pentyl Acetate, followed by a 5s immersion in Xylene and a final 1 minute IPA rinse. On the developed resist, we sputter $\sim$90~\si{\nano\meter} of MoRe, followed by the same lift-off procedure as for the first step. The re-flowed PMGI layer is not removed during by the Anisole during the lift-off and will be removed at a later time.

\textbf{Device Etching}. For the final lithography step, we spin-coat the sample with AR-P 6200-13 at 4000~\si{rpm} followed by a bake at 160~\si{\celsius} for 3 minutes. In this lithography step, we pattern the etchmask for the photonic structures, the acoustic shielding, as well as the boundaries of the piezoelectric resonator. The resist is again developed by immersion for 1 minute in Pentyl Acetate, followed by a 5s immersion in Xylene and a final 1 minute IPA rinse. We transfer the etchmask to the device layer using a reactive-ion etch (RIE) step with a N$\und{2}/$Cl$\und{2}$/BCl$\und{3}$ chemistry with gas flows of 10~SCCM, 20~SCCM, and 10~SCCM, respectively. We etch at a chamber pressure of 6~\si{\micro\bar} for 1:15 minutes. After the RIE step, we remove the remaining etchmask along with the rest of the PMGI layer by immersion in 80~\si{\celsius} dimethyl-formamide (DMF) for 10~minutes, followed by 20~minutes immersion in 90~\si{\celsius} NMP solution. After rinsing in IPA and blow-drying, we spin-coat the sample with S1805 photoresist and use a dicing saw to cut the sample along the optical waveguides (the cut is visible in Fig.~\ref{figure1}b) in order to provide access to the devices with a lensed optical fiber. The resist is then removed by immersion in a 50~\si{\celsius} Acetone bath, which is followed by an IPA rinse.

As a final fabrication step, the sacrificial Al$_{0.64}$Ga$_{0.36}$P layer is selectively removed underneath the GaP device layer using a 1 hour wet etch in a 10\% NH$\und{4}$F solution. The sample is then thoroughly rinsed in water and IPA, and dried in a critical point dryer to avoid unnecessary agitation of the suspended structures.

\subsection{Simulation of the capacitance of the transducer}

\begin{figure}
\includegraphics[width=\columnwidth]{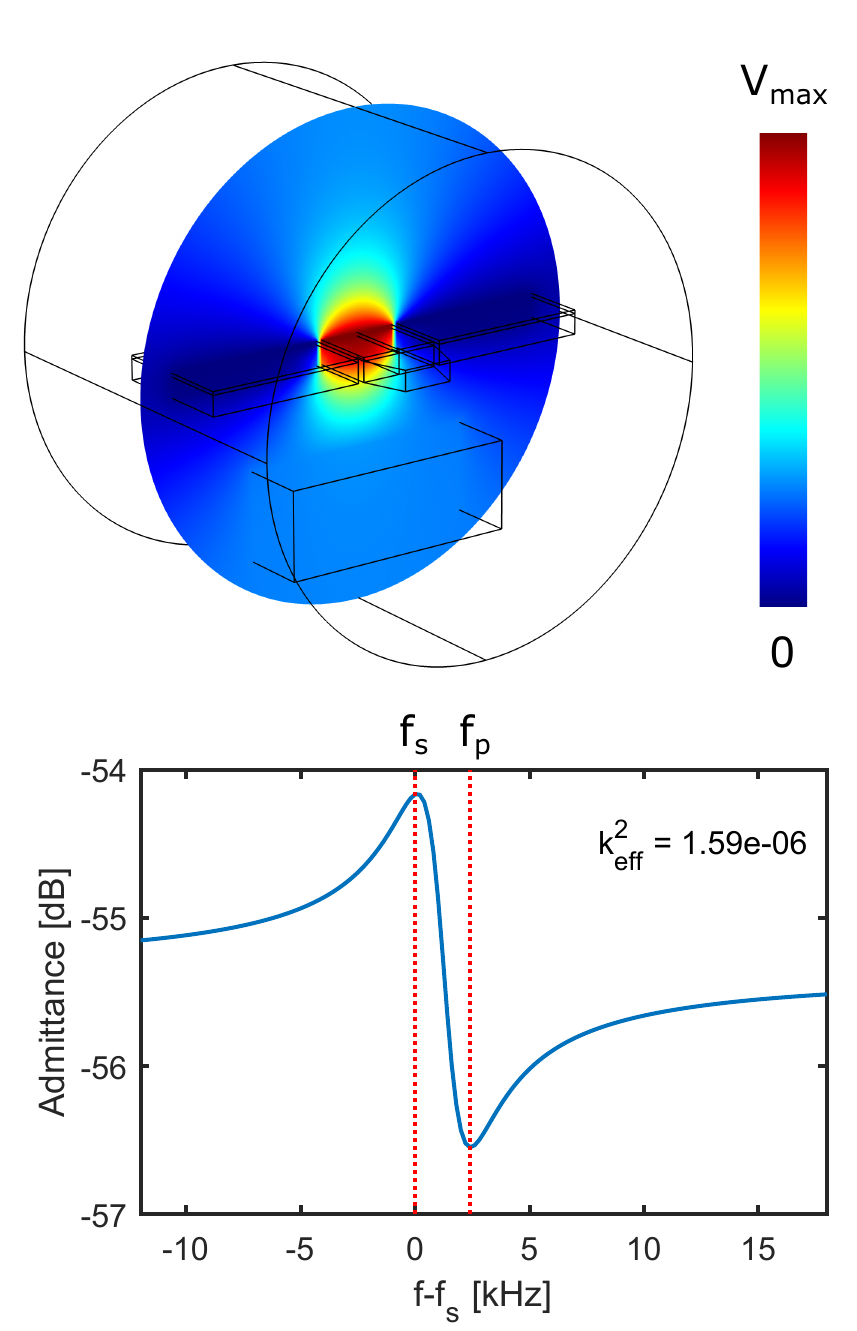}
\caption{\textbf{Electrical Simulations of the Pizeoelectric resonator.} Top:\ stationary electrostatic simulation with a cut plane showing the electric potential. The cut plane is perpendicular to the [110] axis in the crystal. The simulated geometry consists of the piezoelectric resonator (center) with two tapered ends and two side pads. Both the resonator and the pads consist of a 230~\si{\nano\meter}-thick gallium phosphide device layer. On top of the device layer is a 40~\si{\nano\meter} thick film of molybdenum-rhenium (MoRe). The structure is surrounded by an airbox and 1~\si{\micro\meter} underneath, a 1~\si{\micro\meter}-thick block of GaP is placed to account for the effect of the substrate. We extract a capacitance of $C\und{res}$ to be 0.17~\si{\femto\farad}. Bottom:\ Simulated admittance curve for the same structure. We perform a frequency-domain study where we sweep the drive frequency across the breathing mode of the resonator (Fig.~\ref{figure1}d). From the frequencies at which the series ($f\und{s}$) and parallel ($f\und{p}$) resonances occur, we calculate $k\und{eff}^2=1.59\times10^{-6}$.
\label{figureSI1}}
\end{figure}

\begin{figure}
\includegraphics[width=.8\columnwidth]{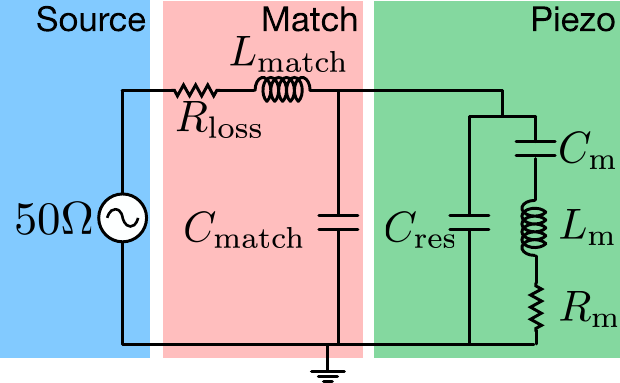}
\caption{\textbf{Piezo-resonator and matching circuit} Circuit representation of the electrical side of the gallium-phosphide transducer. The 50-$\Omega$ source impedance is shown in the blue area on the left. To the right, in the green region, is the piezo-resonator. This is formed of the capacitance between the superconducting plates ($C\und{res}$), in parallel with the effective RLC-circuit formed by the electrically-active mechanical mode. On resonance with the mechanical mode, the circuit dissipates energy into the effective resistor $R\und{m}$. This process can be made more efficient by impedance matching the load to the 50-$\Omega$-source, by using a resonant LC circuit (red region) formed of a high-impedance spiral inductor ($L\und{match}$) and the parasitic capacitance to ground ($C\und{match}$), featuring additional resistive loss $R\und{loss}$.}
\label{figureSIcirc}
\end{figure}

In the Butterworth-van Dyke (BVD) model, the piezoelectric resonator is modeled as a circuit consisting of a capacitance $C\und{res}$, in parallel with a circuit consisting of an effective resistance $R\und{m}$, inductance L$\und{m}$, and capacitance $C\und{m}$ (see Fig. \ref{figureSIcirc}). For our piezoelectric resonator with modes of interest at $\omega\und{m,i}$, the product of $L\und{m}$ and $C\und{m}$ is fixed by $\omega\und{m,i}=1/\sqrt{L\und{m}C\und{m}}$ and the equivalent electrical resistance of the mechanical mode can be expressed as~\cite{Wu2020}:
\begin{equation}
R\und{m} = \frac{\gamma\und{m}}{\omega\und{m,i}^2}\frac{1/k\und{eff}^2-1}{C\und{res}},
\end{equation}
where $k\und{eff}^2$ is the electromechanical coupling coefficient given by
\begin{equation}
k\und{eff}^2 = \frac{C\und{m}}{C\und{m}+C\und{res}}.
\end{equation}
While we can determine $\gamma\und{m}$ and $\omega\und{m,i}$ experimentally, the same is not true for the values of $C\und{res}$ or $k\und{eff}^2$. In order to contextualize the conversion results from the main text and offer potential avenues for improvement, we need to estimate the experimentally inaccessible parameters. We do this by simulating the piezoelectric resonator in COMSOL. In order to obtain the capacitance of the structure, we perform a stationary electrostatic study where we apply 1~\si{\volt} to the central electrode while the side electrodes are grounded (Fig.~\ref{figureSI1}). Then we extract the capacitance and determine $C\und{res}$ to be 0.17~\si{\femto\farad}. In order to extract $k\und{eff}^2$, we perform a frequency domain study of this structure where an AC-voltage oscillating at the drive frequency is applied to the center electrode. We then extract the admittance which is shown in the bottom panel of Figure.~\ref{figureSI1}. We can calculate an estimated $k\und{eff}^2 =(f\und{p}^2-f\und{s}^2)/f\und{p}^2=1.59\times10^{-6}$. Owing to the small expected values of both $k\und{eff}^2$ and $C\und{res}$, we expect the electrical resistance of the mechanical mode, $R\und{m}$ to be large ($2.1$\si{\mega\ohm} for a mechanical quality factor of $1\times10^5$), such that it is inefficiently excited from at 50-$\Omega$ source impedance. To help overcome this expected impedance mismatch, we include a superconducting impedance matching resonator (red area in Fig.~\ref{figureSIcirc}). More details can be found in the Supplementary Information section~\ref{PE_int}.

\subsection{OMC-Waveguide coupling}

\begin{figure}
\includegraphics[width=.8\columnwidth]{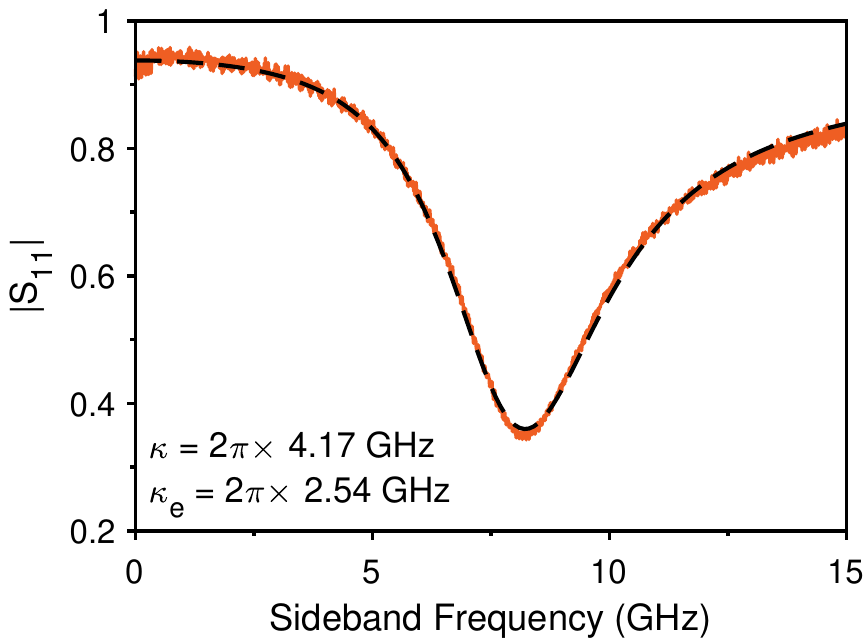}
\caption{\textbf{Characterization of the Optical Cavity.} The orange curve is the reflection parameter from the optical cavity. We stabilize a laser 8-GHz blue-detuned from the optical resonance and scan GHz-frequency sidebands across the optical resonance. For an over-coupled device the reflected light from the two sidebands interfere destructively and the value of $|S\und{11}|$ dips below 0.5. The dashed black curve is a fit to the data.}
\label{figureSI2}
\end{figure}

We access the optical photons in our transducer through an evanescently coupled waveguide (see Fig. \ref{figure1}c of the main manuscript). The optical resonator is coupled to the waveguide at a rate $\kappa\und{e}$, and features an intrinsic loss rate $\kappa\und{i}$, resulting in a total loss rate $\kappa = \kappa\und{e}+\kappa\und{i}$ and limiting the interface efficiency to the overcoupling ratio: $\eta_o = \kappa\und{e}/\left(\kappa\und{e} + \kappa\und{i}\right)$. We can directly access these rates through monitoring the phase of resonant light reflected from the cavity. To do this, we set a laser tone at a fixed detuning a few Gigahertz from the optical resonance, and scan the frequency of a microwave tone we send to an amplitude EOM. The tone results in our detuned carrier tone, $E_0$ and two in-phase sidebands, $E_{\pm}$, of which $E_-$ is resonant with the cavity. Each of these tones is reflected from the cavity with coefficient $r$, given by:
\begin{equation}
r =1-\frac{\kappa\und{e}}{\kappa/2 - 2i\Delta},
\label{ref_E}
\end{equation}
where  $\Delta = \omega\und{l} - \omega\und{c}$ is the detuning of the particular tone at frequency $\omega\und{l}$ from the cavity resonance at  $\omega\und{c}$. We detect the components of the reflected signal at the same frequency as our microwave drive resulting in the complex scattering parameter:
\begin{equation}
S_{11} = E_0E_+^* + E_0^* E_+ + E_0E_-^* + E_0^*E_-,
\label{S11}
\end{equation}
with each tone being reflected by its own value of $r$. Figure \ref{figureSI2} shows the calibrated value of $|S_{11}|$ for the device measured in the main text. We fit the data according to equations \ref{ref_E} and \ref{S11}, with the intrinsic loss rate, the external coupling rate and the exact detuning of the carrier tone $E_0$ as free parameters. We recover a total cavity linewidth of $\kappa = 2\pi\times 4.17$~\si{\giga\hertz}, and an external coupling rate $\kappa\und{e} = 2\pi\times2.54$~\si{\giga\hertz}, from which we extract an optical interface efficiency of $\eta\und{o} = 0.61$, which is the outcoupling efficiency from the optical resonator to the nearby waveguide.

\subsection{Scattering probabilities}

\begin{figure}
\includegraphics[width=1\columnwidth]{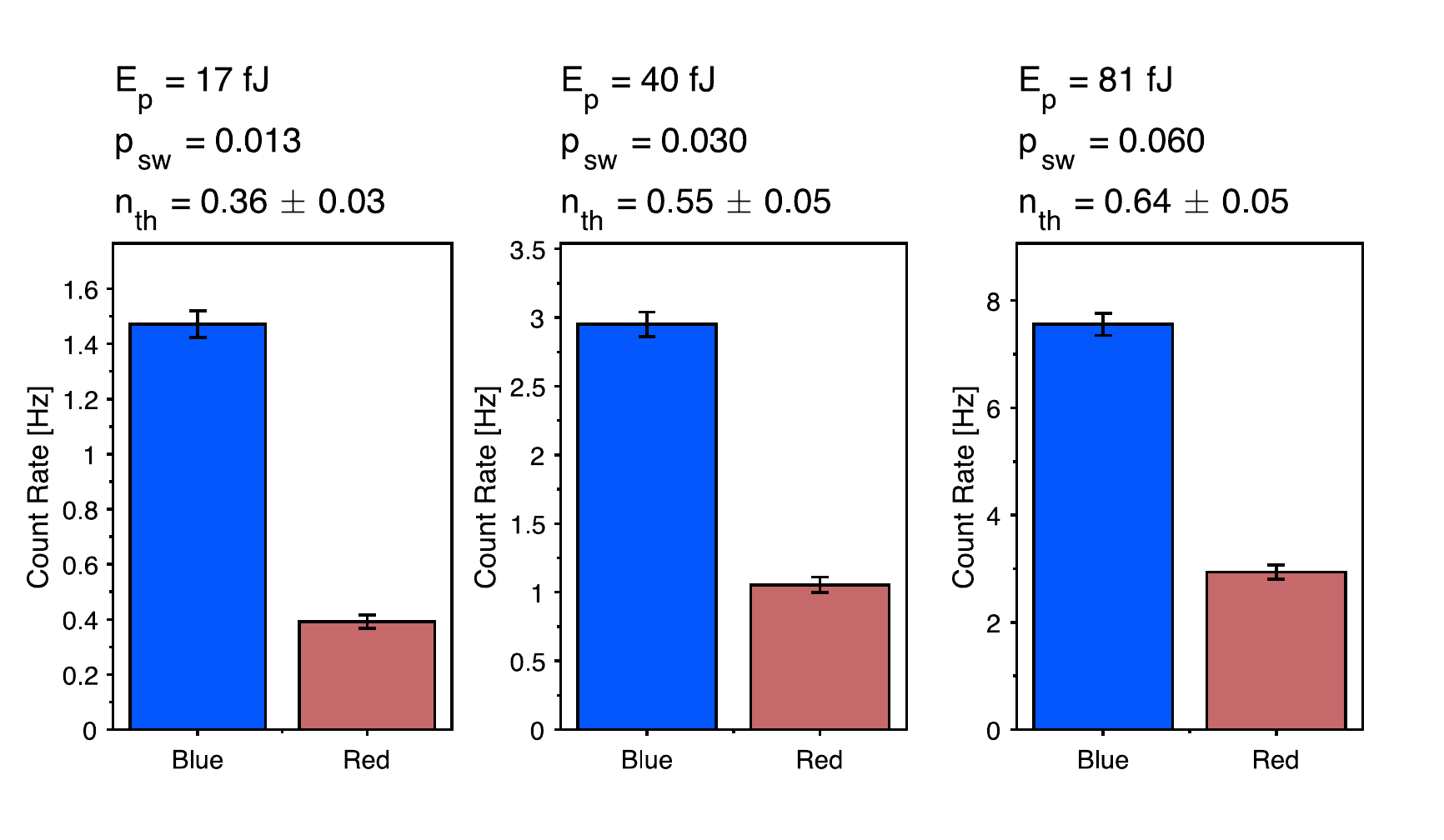}
\caption{\textbf{Sideband Asymmetry.} Recorded count rates for scattering from the blue and red sidebands of the optical device for different pulse energies. The scattering probability $p\und{sw}$ is calculated for the mode at 2.799~GHz. For higher pulse energies, the scattering probability increases, but with increasing thermal noise due to absorption. Error bars are s.d.}
\label{figureSI4}
\end{figure}

In order to examine the different contributions to the overall device efficiency we need to estimate the optomechanical efficiency of the interface $\eta\und{om} = p\und{sw}\eta\und{o}$, where $p\und{sw}$ is the probability of converting a phonon to an optical-frequency photon in the resonator, and $\eta\und{o} = 0.61$. We can estimate the scattering probability, $p\und{sw}$ from~\cite{Hong2017}:
\begin{equation}
p\und{sw}=1-\exp\left(\frac{-4\eta\und{o}g\und{0}^2E\und{p}}{\hbar\omega\und{l}\left(\omega\und{m}^2+\left(\kappa /2\right)^2\right)}\right)	,
\end{equation}
where $g_0$ is the single photon optomechanical coupling strength, $E\und{p}$ the red-detuned pulse energy at frequency $\omega\und{l} = \omega\und{c} - \omega\und{m}$ and $\omega\und{m}$ the frequency of the mechanical mode. 

In order to calibrate the optical pulse energy we measure the coupling of our optical fiber to the device waveguide from the reflection of probe light $100$~\si{\giga\hertz} detuned from the optical resonance. We measure this value to be $\eta\und{coup} = 0.50$. We can estimate the value of $g_0$ from the power-dependent damping rate displayed in Figure~\ref{figure2}d. We fit the FWHM of the conversion peak and use the following expression for the power dependent optomechanical damping rate~\cite{Aspelmeyer2014}: 
\begin{equation}
\gamma\und{m} = \gamma\und{m,0} + n\und{c}g_0^2\left(\mathcal{L}_+ - \mathcal{L}_-\right),
\end{equation}
where $\Delta$ is the detuning of the optical pump from the cavity resonance, $\omega\und{m}$ and $\gamma\und{m,0}$ is the zero-power linewidth, and $\mathcal{L}_{\pm}$ is given by:
\begin{equation}
\mathcal{L}_{\pm} = \frac{\kappa}{\kappa^2/4 + \left(\Delta \pm \omega\und{m}\right)^2} .
\end{equation}
For the mode at 2.799 GHz, we find this value to be  $g_0 = 2\pi\times 700 \pm 8$~\si{\kilo\hertz}, while for the mode at 2.790 GHz, we fit a lower value of $2\pi\times272\pm14$~\si{\kilo\hertz}. 

The scattering probability increases for larger optical pulse powers, however results in an increased thermal population of the mechanical mode, due to optical absorption. Fig. \ref{figureSI4} displays the measured sideband asymmetry for three optical pulse powers, as well as the extracted thermal population, and the scattering probability from the 2.799~\si{\giga\hertz} mode. We use a power of 40~\si{\femto\joule} in the main manuscript, corresponding to a thermal population of $n\und{th} = 0.55\pm0.05$ and a scattering probability of 0.03.

It also is important to note that our device is not fully in the resolved sideband regime ($\omega\und{m}/\kappa = 0.67$), and as a result there is a non-negligible probability for Stokes-scattering from a red-detuned optical pulse. The amount of undesired Stokes scattering from our device can be estimated to be $0.12\times p\und{sw}$~\cite{Wu2020}. In the pulsed protocol used in Figure~\ref{figure3}, we only measure scattered photons at the cavity resonance, selecting the $89\%$ of scattering events that result from the beam-splitter interaction. One consequence of imperfect sideband resolution is amplification-induced noise from the non-supressed two-mode-squeezing interaction. We note that this noise source is limited to a value of $0.12\times p\und{sw}$, which for the pulse energies used in Figure~\ref{figure3} is $<0.01$ phonons, and can be neglected in comparison to absorption-induced noise.

\begin{figure}
\includegraphics[width=.99\columnwidth]{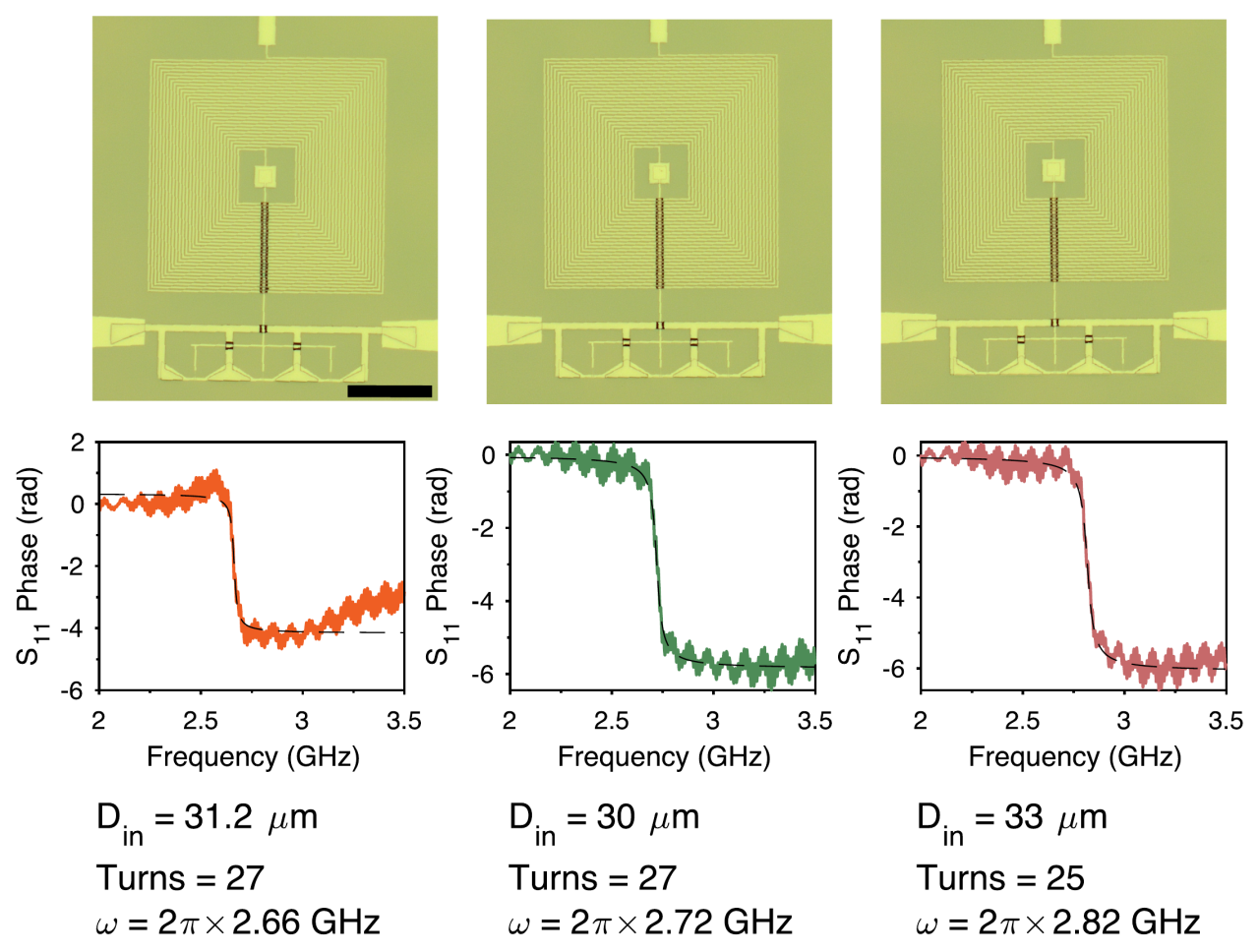}
\caption{\textbf{Inductor Characterization}. Top row:\ Microscope images of three test inductors aimed at targeting the mechanical resonance frequency. The scale bar in the left image is 200~\si{\micro\metre}. Bottom row:\ Phase response of the test inductors measured at 4~K. The text labels show the inner diameter of the inductor, the number of turns and fitted resonance frequency of the devices. Note, due to technical reasons the phase response for the device on the left is slightly reduced.}
\label{figureSI3}
\end{figure}

\subsection{Piezoelectric interface} \label{PE_int}

\begin{figure}
\includegraphics[width=1\columnwidth]{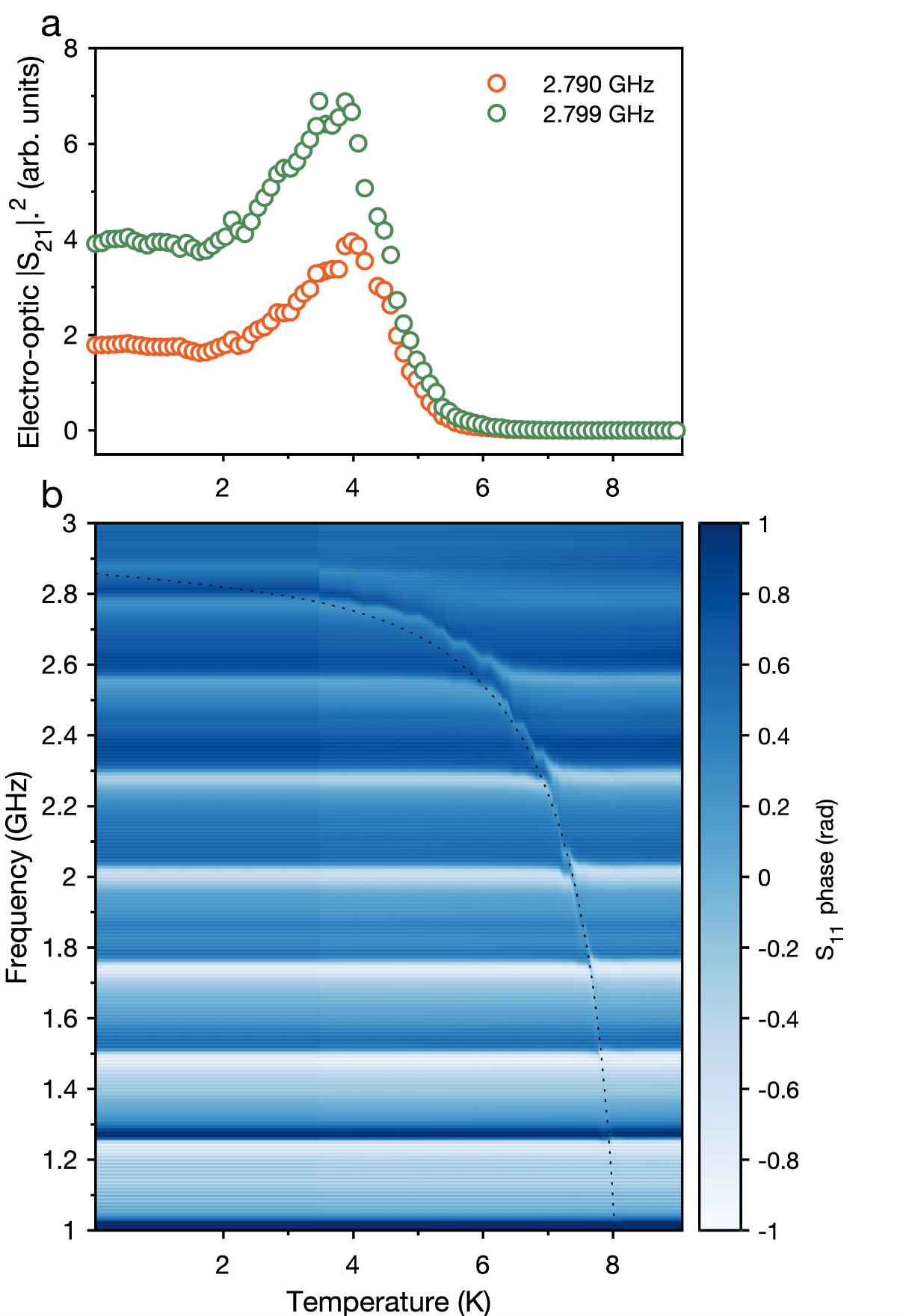}
\caption{\textbf{Electro-Mechanical Interface.} \textbf{a} The evolution in the electro-optic scattering parameter as the device temperature is increased from 20~mK up to 9~K. Displayed are the signal sizes for the modes at 2.790 and 2.799~GHz. \textbf{b} The phase of the reflected signal from the microwave port of the device for the same temperature range. This data was recorded simultaneously with the electro-optic data in panel a. The dotted curve is a fit to the impedance matching circuit resonance frequency, from the BCS-theory of kinetic inductance as the film approaches the critical temperature.}
\label{figureSI5}
\end{figure}

We construct our resonant impedance matching circuit from a thin-film spiral inductor, with the stray capacitance to ground forming the LC-circuit shown in Fig.~\ref{figureSIcirc}. We deposit a film of 90-nm thick MoRe in a square spiral, with track width and spacing of 1~\si{\micro\meter}. We use airbridge-crossovers to route the signal out from the center of the spiral. For the device in the main text, we use a 27-turn inductor with an inner diameter of 30~\si{\micro\meter}.

Figure~\ref{figureSI3} shows microscope images and phase-response from three test inductors fabricated on bulk gallium phosphide, tested at 4~K. The devices show overcoupled resonances close to the target mechanical frequency. Fitting the response of the standard (50-$\Omega$-coupled) RLC circuit reveals a capacitance of $\approx 19$~\si{\femto\farad}, with an inductance of 180~\si{\nano\henry}. The capacitance is slightly larger than the simulated limit of these devices at 12.6~\si{\femto\farad}. The capacitance of these circuits could be further reduced by reducing the size of the track width and film thickness, however this optimization requires further fabrication development to achieve a reliable fabrication process. The impedance matching resonators are directly connected to the 50-$\Omega$ input line, resulting in overcoupled resonators with quality factors of 62, and impedances of $Z\und{match} = \sqrt{L/C} = 3.1$~k$\Omega$. We find that any additional resistive losses $R\und{loss}$ amount to only $2-4\Omega$, small enough to be neglected in comparison to the 50-$\Omega$ source impedance.

We expect the thin-film devices at Millikelvin temperatures to show higher-frequency resonances by approximately 90~MHz, with a blue shift of 30 MHz due to the lower dielectric constant of the AlGaP sacrificial layer~\cite{Robert2013} and a blue-shift of 60~MHz due to the reduced kinetic inductance at lower temperature.

Figure~\ref{figureSI5} displays the result of a temperature sweep of the device, performed by raising the temperature of the dilution fridge cold-plate, during which we continuously measure the electro-optic signal, in the same style as in Figure~\ref{figure2}, as well as the reflected microwave signal. The microwave resonator is partially obscured by reflections in the microwave line, however as the temperature is increased to 8~K, we observe a red-shift of the resonance frequency, consistent with an increasing kinetic inductance in the MoRe film as we approach the critical temperature. The dotted curve in Figure~\ref{figureSI5}b displays a fit of the resonance frequency according to the BCS-theory of the change in kinetic inductance near the critical temperature~\cite{Annunziata2010}, which reproduces the reduction in resonance frequency we observe.

One effect of the increase in film-kinetic inductance is the change of detuning between the mechanical-mode frequencies and the impedance matching resonance. Figure~\ref{figureSI5}a displays the electro-optic signal measured for the mechanical modes at 2.790 and 2.799~GHz through the temperature sweep. We record an increase in the efficiency (proportional to $|S_{21}|^2$) of the transducer as the electrical resonance passes the two mechanical modes at around 4~K. Above this temperature the impedance matching resonance rapidly decreases below 2.79~GHz and the efficiency reduces.

\begin{figure*}
	\includegraphics[width=1.2\columnwidth]{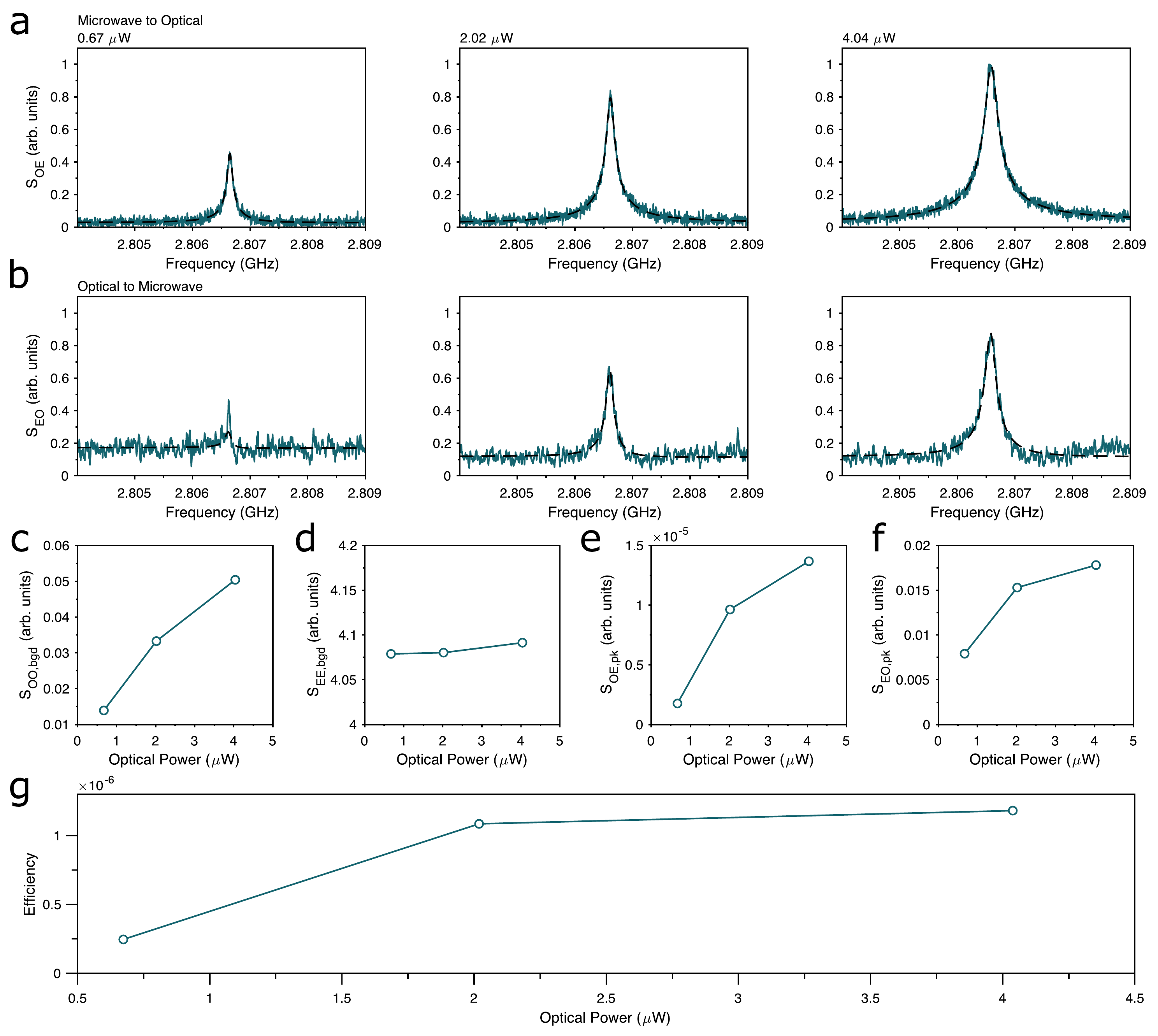}
	\caption{\textbf{Bi-directional transduction in gallium phosphide.} Top Row:\ Microwave-to-optical transduction measured for varying optical input power. Second Row:\ Optical-to-microwave transduction measured for the same device under the same optical pump powers. Third Row:\ Background and extracted peak values for optical and microwave reflection, upconversion and downconversion. Bottom Row:\ Estimated Conversion efficiency.}
	\label{figureSI6}
\end{figure*}

\subsection{Bi-directional Conversion}

The linear conversion process between the microwave and optical domains with our platform is bi-directional, as can be seen in Fig.~\ref{figureSI6}. Here, we also observe down-conversion from optical to microwave frequencies in an analogous way to the up-conversion shown in Figure 2 of the main text. We generate a signal to be down-converted by modulating our red-detuned pump laser with an electro-optic modulator at around 3~GHz. We then directly detect the output microwave signal from the transducer using a cryogenically cooled HEMT amplifier at the 4-K stage of the dilution fridge. The results for both up-conversion and down-conversion for varying optical pump power are displayed in Figure~\ref{figureSI6}. Due to technical reasons, the data were taken using a different device on the same chip as the device in the main manuscript which is in principle identical, except for featuring an inductor with 26 turns and an internal diameter of 31.5~$\mu$m, and featuring a peak in transduction efficiency at 2.807~GHz. While up-conversion and down-conversion occur with the same efficiency, the reduced signal-to-noise in the down-conversion stems from the smaller efficiency in the microwave-frequency detection path.

With bi-directional conversion we can calibrate the losses in the microwave and optical lines to estimate the conversion efficiency. In order to do this, we extract the off-resonant reflection of optical and microwave tones from the device, to provide $S\und{OO, bgd}$ and $S\und{EE,bgd}$, respectively. We fit the transduced fields with square-root Lorentzian profiles, following the expected transduction profile. The fitted amplitudes for up-conversion and down-conversion ($S\und{OE, pk}$  and $S\und{EO, pk}$, respectively) are plotted in Fig.~\ref{figureSI6}e-f. From these four values we can estimate the efficiency by

\begin{equation}
\eta = \frac{S\und{OE, pk}S\und{EO, pk}}{S\und{OO, bgd}S\und{EE, bgd}}.
\end{equation}

The extracted efficiency estimates for the three measured pump-powers are displayed in Fig.~\ref{figureSI6}g. These values show good agreement with pulsed efficiencies extracted for this device, for which we measure an electrical-to-mechanical per-photon efficiency of $3.4\times10^{-6}$.

\end{document}